\documentstyle[aps,manuscript]{revtex}

\begin{document}
\draft
\title{Decay of spin polarized hot carrier current in a quasi one-dimensional 
spin valve structure}
\author{S. Pramanik and S. Bandyopadhyay$\thanks{Corresponding author. e-mail: 
sbandy@vcu.edu}$}
\address{Department of Electrical Engineering,
Virginia Commonwealth University,
Richmond, Virginia 23284}
\author{M. Cahay}
\address{Department of Electrical and Computer Engineering and Computer Science,
University of Cincinnati, Cincinnati, Ohio 45221}

\maketitle

\begin{abstract}
\noindent We study the spatial decay of spin polarized hot carrier current in a  
spin-valve structure consisting of a  semiconductor quantum wire flanked by 
half-metallic ferromagnetic contacts.  The current decays because of  
D'yakonov-Perel' spin relaxation in the semiconductor caused by Rashba spin 
orbit interaction. The associated relaxation 
length is found to decrease with increasing lattice temperature (in the range 
30-77 K) and exhibit a non-monotonic dependence on the electric field driving 
the current.
The relaxation lengths  are several tens of microns which are at least an order 
of magnitude larger than what has been theoretically calculated for 
two-dimensional structures at comparable temperatures, Rashba interaction 
strengths and electric fields. This 
improvement is a consequence of one-dimensional carrier confinement that does 
not necessarily suppress carrier scattering, but nevertheless suppresses 
D'yakonov-Perel' spin relaxation.
\end{abstract}


\pagebreak

Ever since the discovery of the spin-valve effect \cite{dieny}, there has been 
considerable interest in studying spin transport in non-magnetic materials
in which spin polarized carriers are injected from a ferromagnetic contact and 
detected by another ferromagnetic contact. The spin valve structure has 
also been employed to devise novel spintronic devices, such as the so-called 
spin field effect transistor \cite{datta}, where an electron's spin (rather than 
its
charge) is employed to elicit transistor action.

The basic spin valve geometry is shown in the top panel of Fig. 1. It consists 
of a semiconductor channel (assumed to be quasi one-dimensional for this study) 
flanked by two half-metallic ferromagnetic contacts. One contact (called the 
``source'') injects spin polarized current into the channel and thus acts as a 
``spin polarizer''. The other contact acts as a ``spin analyzer'' and is termed 
the ``drain''. Carriers drift from the source to the drain under the influence 
of a driving electric field. When they arrive at the drain, they are transmitted 
 with a probability  $|T|^2$ = $cos^2 (\theta/2)$ where $\theta$ is the angle 
 between the electron's spin polarization at the drain end  and the drain's 
magnetization \cite{datta}. With 
increasing
 degree of spin depolarization in the channel (caused by spin relaxation), the 
average ``misalignment angle'' $\theta$ (for the electron ensemble)
increases and consequently the transmitted current  decreases.  Ultimately, when 
there is no residual spin polarization in the current (i.e. carriers are equally 
likely to have their spins aligned parallel or anti-parallel to the drain's 
magnetization), the transmitted current will fall to 50\% of its maximum value. 
We are interested 
in finding how the (transmitted)
spin polarized current 
falls off with distance along the channel at different driving electric fields 
and temperatures.

Spins depolarize in the channel primarily because of spin-orbit interactions 
caused by bulk 
inversion asymmetry 
(Dresselhaus
spin-orbit coupling) \cite{dresselhaus} and structural inversion asymmetry 
(Rashba spin-orbit 
coupling) \cite{rashba}. These spin-orbit couplings are momentum dependent, and 
because different electrons  have different momenta that change randomly due to 
scattering, the spins become randomized by scattering and the {\it ensemble 
averaged} spin and spin polarized current decay with distance. This mechanism of 
spin relaxation is the D'yakonov-Perel' mechanism \cite{dyakonov} which 
is overwhelmingly dominant in quasi one-dimensional structures over the 
Elliott-Yafet \cite{elliott} or Bir-Aronov-Pikus \cite{bir} mechanisms. The 
spatial decay of spin due to D'yakonov-Perel' mechanism was studied in the past 
by Bournel, et. al. \cite{bournel} and Saikin, et. al. \cite{saikin}
in two-dimensional channels. They mostly dealt with low driving electric fields 
so that transport is linear or quasi-linear. In contrast, 
we have studied the spatial decay in quasi one-dimensional structures of {\it 
both} spin and spin polarized current at high driving electric fields of 1-10 
kV/cm, which result in hot carrier 
transport and non-linear effects.

In a one-dimensional structure, the  spin polarized current due to one electron 
is proportional to  $q v_x |T|^2$ where 
$v_x$ is the ensemble averaged velocity of the electrons along the channel. As 
stated before, the quantity $|T|^2$ 
depends on the 
component of the electron's spin polarization along the magnetization of the 
drain. We will assume that the source and drain are both magnetized along the 
channel's axis (x-axis). This results in the initial spin orientation to be 
along the channel axis. Accordingly,
\begin{eqnarray}
|T|^2 = cos^2 (\theta/2) \nonumber \\
cos(\theta) = S_x/\sqrt{S_x^2 + S_y^2 + S_z^2} = \bar{S_x}~,
\label{transmission}
\end{eqnarray}
where $S_n$ is the spin component along the $n$-axis and $\bar{S_x}$ is the 
normalized value of $S_x$. 

The ensemble averaged  spin polarized current at any position $x$ is  given by
\begin{equation}
I_s (x) = q\sum_{v_x, \bar{S_x}} f(v_x, \bar{S_x}, x) v_x |T(\bar{S_x})|^2 ~,
\label{current}
\end{equation}
where the velocity ($v_x$)- and spin ($\bar{S_x}$)- dependent distribution 
function
$f(v_x, \bar{S_x}, x)$ at any position $x$ is found directly from 
the Monte Carlo simulator described in ref. \cite{sandipan_prb} (all pertinent 
details of the simulator can be found in ref. \cite{sandipan_prb} and will not 
be repeated here). We only mention that in the simulator, we use a parabolic 
energy versus velocity dispersion relation $E = (\hbar^2/2m^*)(n \pi/W_z)^2 + 
(\hbar^2/2m^*)( \pi/W_y)^2 + (1/2)m^* v_x^2$ ($n$ is the subband index in the 
z-direction), neglecting any band structure non-parabolicity which is not 
important in the energy ranges encountered \cite{bandy}. This 
dispersion relation allows us to calculate the velocity $v_x$ from the carrier 
energy $E$ and subband index $n$ (which are tracked in the simulator) very 
easily. If instead we used the energy versus wavevector relation (which is 
traditional) and then attempted to find $v_x$ from the velocity versus 
wavevector relation, it would have been immensely complicated. The reason is 
that the velocity (or energy) versus wavevector relation is {\it spin-dependent} 
in the presence of the Rashba effect \cite{molenkamp} and becomes even more 
complicated if the Rashba effect is strong which leads to spin mixing effects 
\cite{governale}. These complications would have been overwhelming in our case 
since we have a continuous distribution of spin and hence would have been faced 
with a denumerably infinite number of energy versus wavevector relations. The 
way to
avoid this daunting complication (and the associated numerical cost) 
is to use the energy-velocity relation which is {\it spin-independent} instead 
of the energy-wavevector relation which is {\it spin-dependent}. 

In the simulation, carriers are injected into a quasi  one-dimensional 
GaAs channel of rectangular cross section 30 nm $\times$ 4 nm (see top panel of 
Fig. 1). We have assumed that there is a transverse electric 
field of 100 kV/cm (in the y-direction) that gives rise to a structural 
inversion asymmetry and induces a Rashba effect in the channel. This field 
perturbs the subband energies in the channel but only slightly. The transverse 
voltage drop over a 4 nm wide channel due to this field is 40 meV while the 
lowest subband energy is 355 meV. Therefore, the perturbation is 11\% for the 
lowest subband and progressively decreases for higher subbands. Consequently, we 
neglect this perturbation. Electrons are injected from a Fermi Dirac 
distribution with their spins all 
aligned along the channel axis (x-axis) in order to simulate the spin polarizer. 
 At any 
given position $x$, we find the spin vector $\bar{S_x}$ and compute the quantity 
$|T(\bar{S_x})|^2$ for every electron
using Equation (\ref{transmission}).
We also find the velocity $v_x$ for every electron at position $x$ and then 
compute the spin polarized current $I_s$ by performing the ensemble averaging 
given by Equation (\ref{current}). We have found $I_s$ versus position $x$ for 
four different channel 
electric 
fields of 1, 2, 4 and 10 kV/cm and two different temperatures of 30 and 77 K.

In the Monte Carlo simulation, we find that typically the lowest 4 subbands are 
occupied by electrons. The Dresselhaus interaction strength is different in 
different subbands because the interaction strength is proportional to $(n 
\pi/W)^2$ where $W$ is the transverse dimension (4 nm in our case) and $n$ is 
the subband index. As electrons are scattered between various subbands, they 
witness varying spin orbit interaction which results in a decaying envelope of 
the {\it ensemble averaged} spin (or spin polarized current). This is the cause 
of D'yakonov-Perel relaxation in quasi 1-d structures.

In Fig. 1, we show the spatial decay of the normalized spin polarized current 
$I_s$   for the four 
different (x-directed) channel electric fields at a temperature of 30 K. In Fig. 
2, we show the same quantity (along with the spatial decay of the ensemble 
averaged spin component $\bar{S_x}$) at an electric field of 2 kV/cm at 
temperatures of 30 and 77 K. Spin depolarization is complete when  $I_s$ reaches 
a value of 0.5. At this point, an electron  is equally likely  to have its spin 
aligned parallel or anti-parallel to the drain's magnetization (and therefore it 
is equally likely to be transmitted or blocked).
   We can define a ``relaxation length'' as the distance 
over which the injected spin polarized current decays to 50\% of its value (i.e. 
becomes completely depolarized). Table I 
gives the relaxation lengths at different electric field strengths and different 
temperatures.

As expected, the relaxation length decreases with increasing carrier temperature 
because of  increased scattering that causes increased spin depolarization. The 
electric 
field, on the other hand, has two opposing effects. The scattering rate 
increases slowly with the electric field, but so does the ensemble averaged 
carrier drift velocity until the saturation velocity is reached. A larger drift 
velocity makes the carriers travel a greater distance before getting 
depolarized. Consequently, the relaxation length at first increases with 
increasing electric field, but once the drift velocity begins to saturate, the 
increased scattering  takes over
and the relaxation length starts to decrease with increasing electric field.
The dependence of relaxation length on the electric field is therefore  
non-monotonic.

Based on the data in Table I, we find that the relaxation length for spin 
polarized current  is very large (between 20 and 100 $\mu$m for the cases 
considered). This is at least an order of magnitude larger than what was 
calculated for two-dimensional structures \cite{bournel,saikin} at comparable 
temperatures and driving electric fields. This difference is not due to any 
suppression of scattering. In fact, even though elastic scattering is suppressed 
in quasi one-dimensional structures \cite{sakaki}, inelastic scattering is not 
\cite{telang}, and the calculated mobility in one-dimensional structures in this 
temperature range is less than that in bulk \cite{telang2}. The true origin of 
the difference lies in the fact that Dresselhaus and Rashba interactions cause a 
carrier's spin to precess slowly (during free 
flight) about a so-called ``spin precession vector'' that is defined by the 
carrier's momentum \cite{bournel}. In a one-dimensional structure,  a carrier is 
free to move only along one direction, and therefore the 
Rashba or the Dresselhaus spin precession vector always points along one 
particular direction. Scattering can change its magnitude, but not its 
direction. This leads to slow spin relaxation. In contrast, scattering can 
change both the magnitude and the direction of the spin precession vector in 
two- or three-dimensional structures.
Therefore, spin depolarizes much faster in multi-dimensional structures.

Before concluding this Letter, we should mention that in the type of structures 
considered here, there is always a magnetic field in the channel caused by the 
ferromagnetic contacts. This field, however weak, ensures that the eigenstates 
in the channel are not spin eigenstates \cite{cahay_prl}. Therefore, even 
non-magnetic scatterers can cause  spin relaxation \cite{cahay_prb}. This 
mechanism has not been considered here, since we have not considered the channel 
magnetic field. In the absence of this mechanism, we have shown that spin 
relaxation length of carriers is very 
large in quasi one-dimensional structures, even at elevated temperatures and 
high electric fields. Large spin relaxation lengths have been observed before in 
multi-dimensional structures, but only at low driving electric fields and low 
temperatures \cite{awschalom}. One dimensional confinement can extend the range 
to high electric fields and elevated temperatures, which are required for 
realistic device applications. 

The work of S. P and S. B. were supported by the US National Science Foundation 
under grant ECS-0196554.

\pagebreak

%\bibliographystyle{/d/gady/Style/yzaip}
%\bibliography{/d/rr/latexlib/one}

\begin{references}

\bibitem{dieny}
B. Dieny, V. S. Speriosu, S. S. P. Parkin, B. A. Gurney, D. R. Wilhoit and D. 
Mauri, Phys. Rev. B, {\bf 43}, 1297 (1991).

\bibitem{datta} 
S. Datta and B. Das, Appl. Phys. Lett., {\bf 56}, 665 (1990).

\bibitem{dresselhaus}
G. Dresselhaus, Phys. Rev., {\bf 100}, 580 (1955).

\bibitem{rashba}
E. I. Rashba, Sov. Phys. Semicond., {\bf 2}, 1109 (1960); 
Y. A. Bychkov and E. I. Rashba, J. Phys. C, {\bf 17}, 6039 (1984).

\bibitem{dyakonov}
M. I. D'yakonov and V. I. Perel', Sov. Phys. Solid State, {\bf 13}, 3023 (1972).

\bibitem{elliott}
R. J. Elliott, Phys. Rev., {\bf 96}, 266 (1954).

\bibitem{bir}
G. L. Bir, A. G. Aronov and G. E. Pikus, Sov. Phys. JETP, {\bf 42}, 705 (1976).

\bibitem{bandy}
It is possible to introduce some band structure non-parabolicity by using an 
energy-dependent effective mass. However, this is not important. The carrier 
kinetic energies remain small in every subband so that non-parabolicity effects 
are never significant. If the energy of a carrier in the lowest subbands begins 
to increase, the carrier suffers intersubband transition to a higher subband by 
absorbing or emitting phonons.
This process keeps the kinetic energy in every subband small and the carrier 
temperature remains very close to the lattice temperature. The intersubband 
scattering
(not intervalley transfer) is also responsible for velocity saturation in the 
quantum wire.

\bibitem{molenkamp}
L. W. Molenkamp, G. Schmidt and G. E. W. Bauer, Phys. Rev. B, {\bf 64}, 
121202(R)  (2001).

\bibitem{governale}
M. Governale and U. Z\"ulicke, Phys. Rev. B, {\bf 66}, 073311 (2002).

\bibitem{bournel}
A. Bournel, P. Dollfus, S. Galdin, F-X Musalem and P. Hesto, Solid State
Commun., {\bf 104}, 85 (1997); A. Bournel, P. Dollfus, P. Bruno and P. Hesto, 
Mat. Sci. Forum, {\bf 297-298}, 205 (1999); A. Bournel, V. Delmouly, P. Dollfus, 
G. Tremblay and P. Hesto, Physica E, {\bf 10}, 86 (2001).

\bibitem{saikin}
S. Saikin, M. Shen, M. C. Cheng and V. Privman, 
www.arXiv.org/cond-mat/0212610; M. Shen. S. Saikin, M. C. Cheng and V. Privman, 
www.arXiv.org/cond-mat/0302395.


\bibitem{sandipan_prb}
S. Pramanik, S. Bandyopadhyay and M. Cahay, Phys. Rev. B, {\bf 68}, 075313 
(2003)..


\bibitem{sakaki}
H. Sakaki, Jpn. J. Appl. Phys., {\bf 19}, L735 (1980).

\bibitem{telang}
N. Telang and S. Bandyopadhyay, Phys. Rev. B, {\bf 48}, 18002 (1993).

\bibitem{telang2}
N. Telang and S. Bandyopadhyay, Appl. Phys. Lett., {\bf 66}, 1623 (1995); Phys. 
Rev. B, {\bf 51}, 9728 (1995).

\bibitem{cahay_prl}
M. Cahay and S. Bandyopadhyay, Phys. Rev. B, {\bf 68}, 115316 (2003).


\bibitem{cahay_prb}
M. Cahay and S. Bandyopadhyay, submitted to Phys. Rev. B.


\bibitem{awschalom}
D. D. Awschalom and J. M. Kikkawa, Phys. Today, {\bf 52}, 33 (1999);
Phys. Rev. Lett., {\bf 80}, 4313 (1998); Nature (London), {\bf 397}, 139 (1999).




\end{references}

\pagebreak

\begin{figure}
\caption[]{\small Spatial decay of the normalized spin polarized current  in a  
GaAs quantum wire channel of rectangular 
cross-section 30 nm $\times$ 4 nm.  The results are shown for four different 
channel electric fields of 1, 2, 4 and 10 kV/cm at an electron temperature of 30 
K. The top panel shows schematic of a spin valve with a quasi one-dimensional 
channel.
The half-metallic ferromagnetic source and drain contacts act as spin polarizers 
and analyzers, 
while the gate terminal is used to apply a transverse electric field on the 
channel to induce a
Rashba effect.
}
\end{figure}

\begin{figure}
\caption[]{\small Spatial decay of the normalized spin polarized current and the 
injected spin vector in the channel of 
 Fig. 1. The results are shown for two different 
temperatures of 30 K and 77 K at a channel electric field of  2 kV/cm.
}
\end{figure}

\newpage 

\begin{table}
\caption{Spin relaxation length dependence on temperature and driving electric 
field.}
\bigskip
\begin{tabular}{|c|c|c|}
\hline
Electric field (kV/cm) & Temperature (K) & Spin relaxation ($\mu$m) \\
\hline
& & \\
1.0 & 30 & 20 \\
2.0 & 30 & 60 \\
4.0 & 30 & 100 \\
10.0 & 30 & 50 \\
2.0 & 77 & 30 \\
& & \\
\hline
\end{tabular}
\end{table}

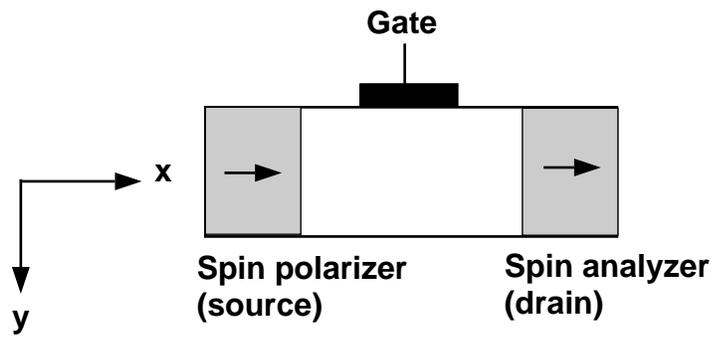
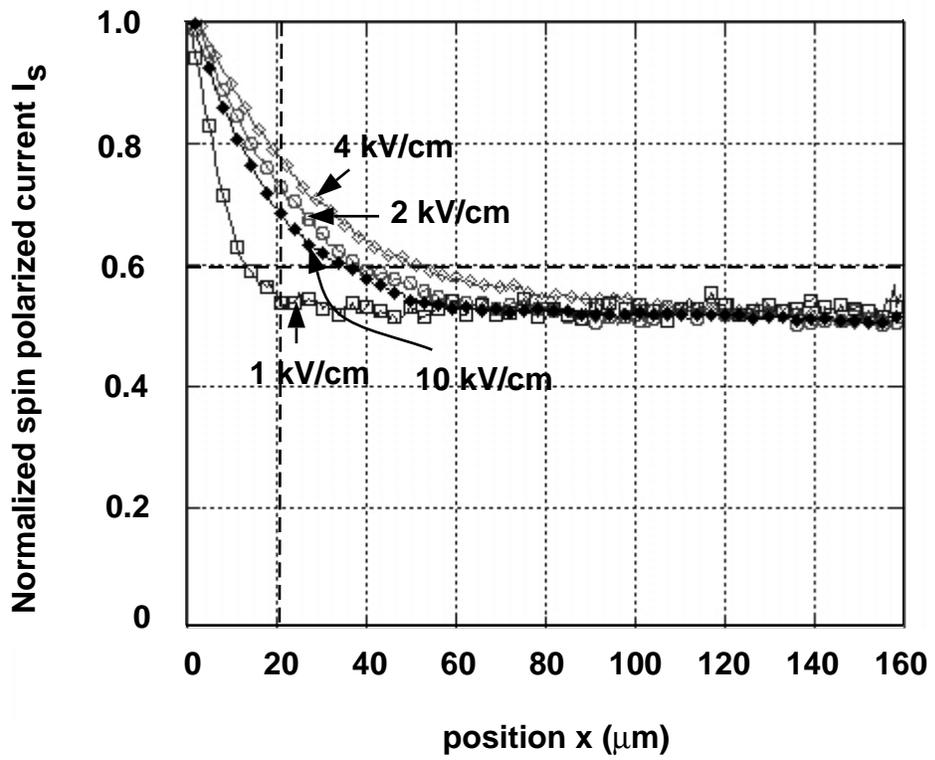
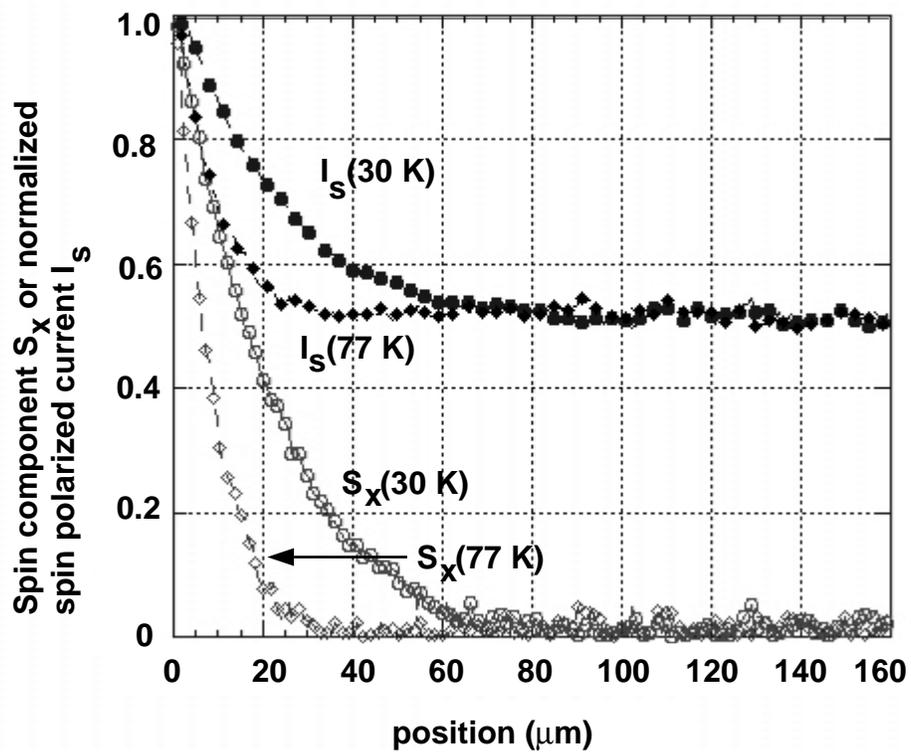
\end{document}